\documentclass[aip,reprint]{revtex4-1}

\usepackage{color}
\usepackage{graphicx}
\usepackage{amsmath}
\usepackage{subcaption}

\usepackage{natbib}

\begin{document}
\title{Resistive Diffusion in Magnetized ICF Implosions: Reduced Magnetic Stabilization of the Richtmyer Meshkov Instability}

\author{C. A. Walsh}
\email{walsh34@llnl.gov}
\affiliation{Lawrence Livermore National Laboratory}
\author{D. J. Strozzi}
\affiliation{Lawrence Livermore National Laboratory}
\author{H. Sio}
\affiliation{Lawrence Livermore National Laboratory}
\author{B. B. Pollock}
\affiliation{Lawrence Livermore National Laboratory}
\author{B. D. Appelbe}
\affiliation{Imperial College London}
\author{A. J. Crilly}
\affiliation{Imperial College London}
\author{S. O'Neill}
\affiliation{Imperial College London}
\author{C. Weber}
\affiliation{Lawrence Livermore National Laboratory}
\author{J. P. Chittenden}
\affiliation{Imperial College London}
\author{J. D. Moody}
\affiliation{Lawrence Livermore National Laboratory}

\date{\today}

\begin{abstract}
	Resistive diffusion is typically regarded to be negligible in magnetized ICF experiments, with magnetic flux effectively compressed during the implosion. In this work the Richtmyer Meshkov instability at the ice-ablator interface is taken as an example for investigating resistive effects. For a high temperature ($\approx$100eV) interface with magnetic field applied perpendicular to shock propagation, perturbation growth is suppressed by magnetic tension. However, for lower temperature interfaces the resistive diffusion prevents substantial magnetic field twisting at small scales. ICF implosion simulations are then used to assess magnetic diffusivity at different interfaces; the ice-ablator interface is found to be too resistive for the magnetic fields to enhance stability. For Rayleigh-Taylor growth at the hot-spot edge, on the other hand, resistivity is estimated to only be a secondary effect, as seen in previous simulation studies.

\end{abstract}
\maketitle

Inertial confinement fusion (ICF) seeks to compress spherical capsules to reach high temperatures and densities in the fusion fuel, with recent experiments reaching the ignition regime \cite{abu-shwareb2022}. The implosions are beset by ubiquitous symmetry issues\cite{clark2019}; the focus here is on the interface between the deuterium-tritium ice and the capsule ablator. It has been conjectured that mixing at this interface due to the Richtmyer Meshkov instability \cite{holmes1999,vetter1995,jacobs1996,zhou2017} reduces the resulting fuel compression \cite{PhysRevLett.117.075002}. The high density carbon (HDC) micro-structure is expected to be the seed for asymmetry \cite{ali2018}, which is then amplified as the implosion shock-waves initiate Richtmyer-Meshkov growth \cite{PhysRevLett.117.075002,Smalyuk_2019,weber2023}. Modes as high as 2000 are expected to degrade capsule performance \cite{weber2023}.

In recent years the National Ignition Facility has added the capability to impose magnetic fields onto implosions \cite{moody2021a}, with experiments demonstrating a 40\% enhancement in ion temperature and 3x increase in neutron yield with a 26T axial field \cite{moody2022,sio2023}. On the OMEGA Laser Facility the enhancements have been 15\% and 30\% \cite{chang2011}. These results are in-line with predicted suppression of thermal conduction losses from the hot-spot \cite{walsh2022}. The applied magnetic fields are also predicted to reduce perturbation growth at the hot-spot edge, where the implosion is Rayleigh-Taylor unstable \cite{walsh2019,walsh2021magnetized}; this is due to tension in the magnetic field lines acting to smooth out the hot-spot surface.

The magnetized Richtmyer-Meshkov instability has been investigated numerically \cite{samtaney2003, sano2013, sano2012} and theoretically \cite{cao2008}, finding substantial reduction in mix at interfaces. Research in this manuscript investigates such a phenomenon in the context of ICF implosions. While suppression is observed for some test cases, realistic interface temperatures are found to be too resistive to support any significant reduction in Richtmyer Meshkov growth rates.

The Richtmyer-Meshkov instability is most easily understood as the impulsive limit of the Rayleigh-Taylor instability; for that reason, a description of the latter instability will be the first focus. A qualitative derivation of the growth rate using simple energetics arguments is preferred over a more comprehensive stability analysis. 

The Rayleigh-Taylor instability occurs when an interface with perturbation wavelength $\lambda$ undergoes an acceleration, $g$. The acceleration is directed from material one to material 2, with densities $\rho_1$ and $\rho_2$ respectively. A spike-bubble height of $h$ corresponds to the spike growing a distance $h/2$; this releases potential energy $(\rho_1-\rho_2) g h/2$. If the system is assumed to be incompressible, the resultant motion is vortical with the dense spike pushing low density material laterally; for small perturbations ($h/\lambda \ll 1$) there is more material moving laterally than vertically by a factor of $\lambda/h$. Therefore, balancing the kinetic and potential energy gives\cite{betti2001}: 
\begin{equation}
\Bigg(\frac{\partial h}{\partial t}\Bigg)_{RTI}^2 = h^2 \frac{ A_t g }{\lambda} \label{eq:RTI}
\end{equation}
Where the Atwood number $A_t$ is defined as $(\rho_1 - \rho_2)/(\rho_1 + \rho_2)$.  Equation \ref{eq:RTI} gives an exponential growth rate of perturbation amplitude $h = h_0 e^{t\sqrt{A_t g/\lambda}}$, where $h_0$ is the initial perturbation height.

The Richtmyer-Meshkov instability is seeded by a shock moving through a perturbed interface, rather than a continuous acceleration. The shock delivers an impulse to the plasma, increasing its velocity by $\Delta V$. Converting this to an acceleration (so that knowledge of the Rayleigh-Taylor instability growth can be used) gives $g=\Delta V / \Delta t$, where $\Delta t$ is the shock transit time through the plasma. 

In order to eliminate $\Delta t$, the second derivative of the Rayleigh-Taylor perturbation height growth is used $\partial^2h/\partial t^2 = h A_t g/\lambda$. For Richtmyer-Meshkov growth, the shock imparts a certain perturbation growth rate during transit and then does not affect the perturbation thereafter; i.e. $h$ is replaced by $h_0$. Integrating gives $\partial h/\partial t = h_0 A_t g\Delta t/\lambda$. Therefore, the following Richtmyer Meshkov growth rate is obtained:

\begin{equation}
\Bigg(\frac{\partial h}{\partial t}\Bigg)_{RM}^2 = \Bigg(\frac{\Delta V A_t h_0}{\lambda}\Bigg)^2 \label{eq:RM}
\end{equation}

Note that if the Atwood number is negative (as at the ice-ablator interface in an ICF implosion) the Rayleigh Taylor growth stabilizes the system. For a typical ICF implosion without magnetic fields, this is thought to be the primary mechanism of stabilization for the Richtmyer Meshkov instability \cite{weber2023}. Combining equations \ref{eq:RM} and \ref{eq:RTI} and setting the growth rate equal to zero gives the spike height at which the Rayleigh Taylor instability saturates the Richtmyer Meshkov growth:

\begin{figure}
	\centering
	\centering
	\includegraphics[scale=0.5]{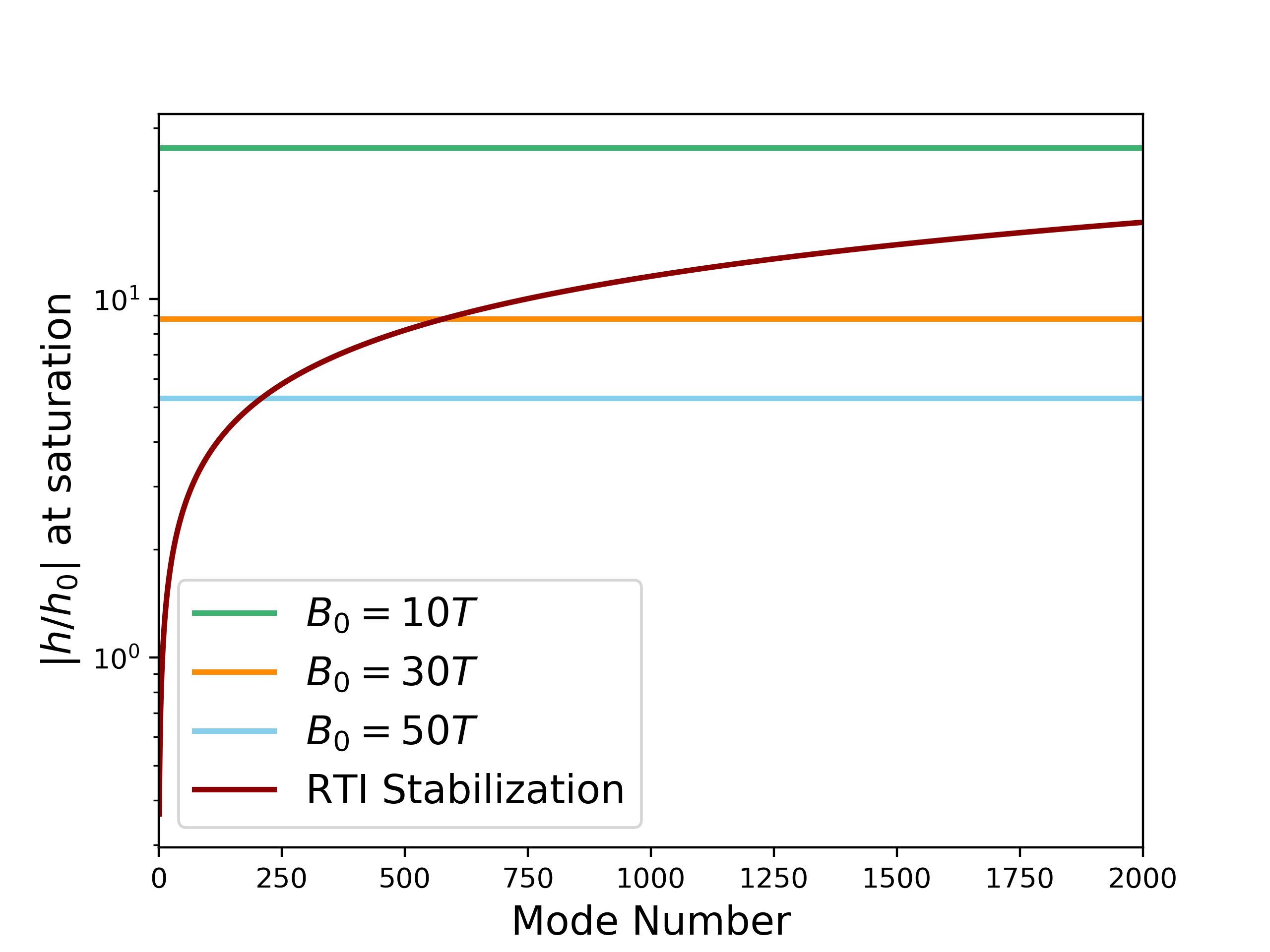} \caption{ \label{fig:h_h0} Stabilization height against perturbation mode number of the linear Richtmyer-Meshkov instability due to the Rayleigh-Taylor instability (RTI) and 3 different transverse magnetic field strengths. }	
\end{figure}

\begin{equation}
	\frac{h_{sat,RTI}}{h_0} =  \pm \Delta V \sqrt{\frac{A_t}{\lambda g}} \label{eq:hsat_RTI}
\end{equation} 

Figure \ref{fig:h_h0} plots equation \ref{eq:hsat_RTI} against mode number for quantities typical to an ICF implosion ($\Delta V =40$km/s, $A_t = -1/3$, $\rho=10$kg/m$^3$, $g=5\times 10^{13}$m/s$^2$ capsule radius $R=$500$\mu$m). As can be seen, RTI stabilization gives the greatest amplitude perturbations for the smallest scales.  

Magnetic fields can modify perturbation growth rates either by directly introducing a stabilizing Lorentz force or through a modification to the thermal transport \cite{walsh2019,walsh2021magnetized}. However, the ablator-ice interface is relatively cold throughout the implosion (as will be discussed in section \ref{sec:postprocess}), meaning that heat-flow is found to be relatively insignificant. The Lorentz force can be reformed into two components: the magnetic pressure ($\nabla (\underline{B} \cdot \underline{B})/2\mu_0$) and the magnetic tension ($\underline{B} \cdot \nabla \underline{B}/\mu_0$). This work finds the magnetic tension to be the significant contributor to modifications to Richtmyer Meshkov growth.

The magnetic tension stabilizes the Richtmyer-Meshkov growth when the magnetic field is in the plane of the interface. As the perturbation grows, it deforms the magnetic field lines; the magnetic tension is a force that acts to straighten those field lines out again. For small amplitude perturbations, the tension force is $\underline{B} \cdot \nabla \underline{B}/\mu_0 \approx |\underline{B}|^2 /\lambda \mu_0$; the energy expended by the plasma onto the magnetic field is this multiplied by $h/2$. Equating this with the kinetic energy shows the stabilizing influence of the magnetic field \cite{manuel2021}:

\begin{equation}
\Bigg(\frac{\partial h}{\partial t}\Bigg)_{B}^2 =  -\frac{|\underline{B}|^2}{\mu_0 (\rho_1 + \rho_2)}\frac{h^2}{\lambda^2} \label{eq:Bstab}
\end{equation}

The saturation height of the Richtmeyer-Meshkov instability due to the magnetic tension can be found by combining \ref{eq:RM} and \ref{eq:Bstab}:

\begin{equation}
\frac{h_{sat,B}}{h_0} =  \pm \Delta V A_t\frac{\sqrt{(\rho_1 + \rho_2) \mu_0}}{|\underline{B}|} \label{eq:hsat_B}
\end{equation} 

The saturation heights due to the applied magnetic field have also been included in figure \ref{fig:h_h0}, with the magnetic fields assumed to have been amplified by convergence in the ice ($|\underline{B}|/B_0 = (\rho/\rho_0)^{2/3}$). An initial 30T magnetic field improves the ice-ablator interface stability for modes greater than 570, while a 50T field improves stability for modes greater than 200. For mode 2000, which is expected to be a detrimental length scale for ICF implosions \cite{weber2023}, the 50T field is expected to reduce the amplitude by a factor of 3. Clearly these basic estimates motivate the more detailed numerical study carried out in this paper. 

The Richtmyer Meshkov simulations in this paper use the extended-magnetohydrodynamics code Gorgon \cite{ciardi2007,chittenden2009,walsh2017}. The magnetic transport in Gorgon includes Nernst advection \cite{walsh2020} and resistive diffusion. The Biermann Battery term is found to be insignificant during the implosion phase of ICF capsules \cite{walsh2021a}, but has been included nonetheless in all simulations. The transport coefficients use a form that has been updated since Epperlein and Haines \cite{epperlein1986,sadler2021,davies2021}, which were found to make a significant difference in capsule implosions \cite{2021}. The magnetized heat-flow is anisotropic, including the Righi-Leduc component \cite{walsh2018a}.

\begin{figure}
	\centering
	\centering
	\includegraphics[scale=0.5]{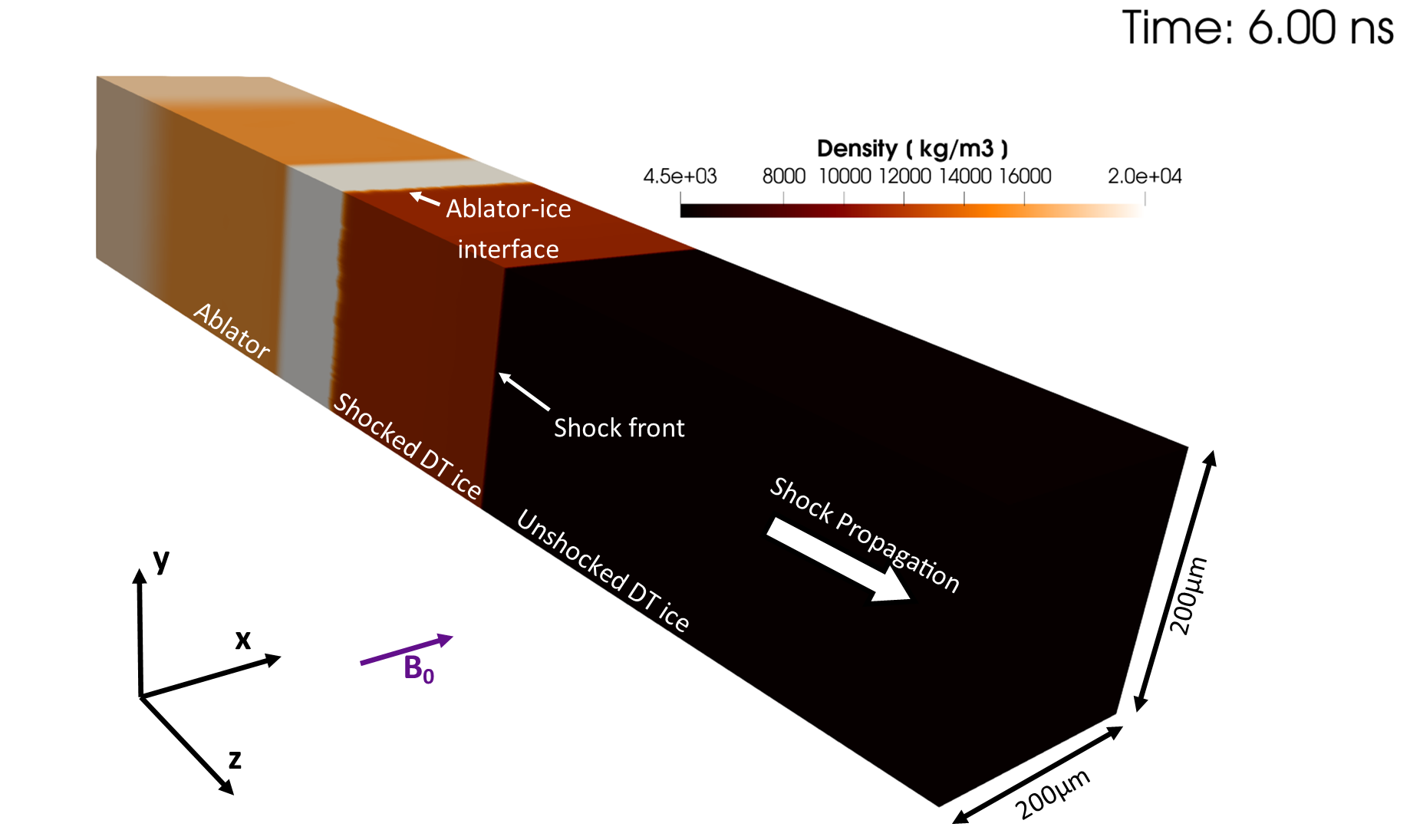} \caption{ \label{fig:3D} Depiction of the 3-D Richtmyer-Meshkov instability test problem used throughout this paper. An impulse is applied to the top left boundary (low $z$) that sends a shock through the system. A multi-mode perturbation is present at the ablator-ice interface. A magnetic field is applied along $x$.   }	
\end{figure}

A test problem is utilized in this study to assess how magnetic fields affect Richtmyer Meshkov growth in 3-D, with figure \ref{fig:3D} used as an aide to orient the reader. A region of compressed HDC ablator is initialized with density $10^4$ kg/m$^3$, adjacent to a compressed DT ice of density $5000$ kg/m$^3$. The temperature at the interface is varied throughout this paper in order to demonstrate the impact of resistive diffusion. Both plasma species begin at rest, with an impulse sending a shock through the system along $z$ going through the ablator first and accelerating the plasma to $40$km/s. These numbers are in-line with what is expected in an ICF implosion \cite{weber2023}.

The simulation box is Cartesian, so any convergence effects are neglected. The boundaries in $x$ and $y$ are periodic. The simulation box is $200\mu m \times 200 \mu m \times 1200 \mu m$ with a resolution of 0.5$\mu m$ in each direction. The mode numbers quoted are adjusted to reflect that the test simulation is only one patch of the overall capsule surface; a capsule radius of $500\mu$m is assumed, although the test problem geometry does not include any convergence effects. 
\begin{figure}
	\centering
	\centering
	\includegraphics[scale=0.55]{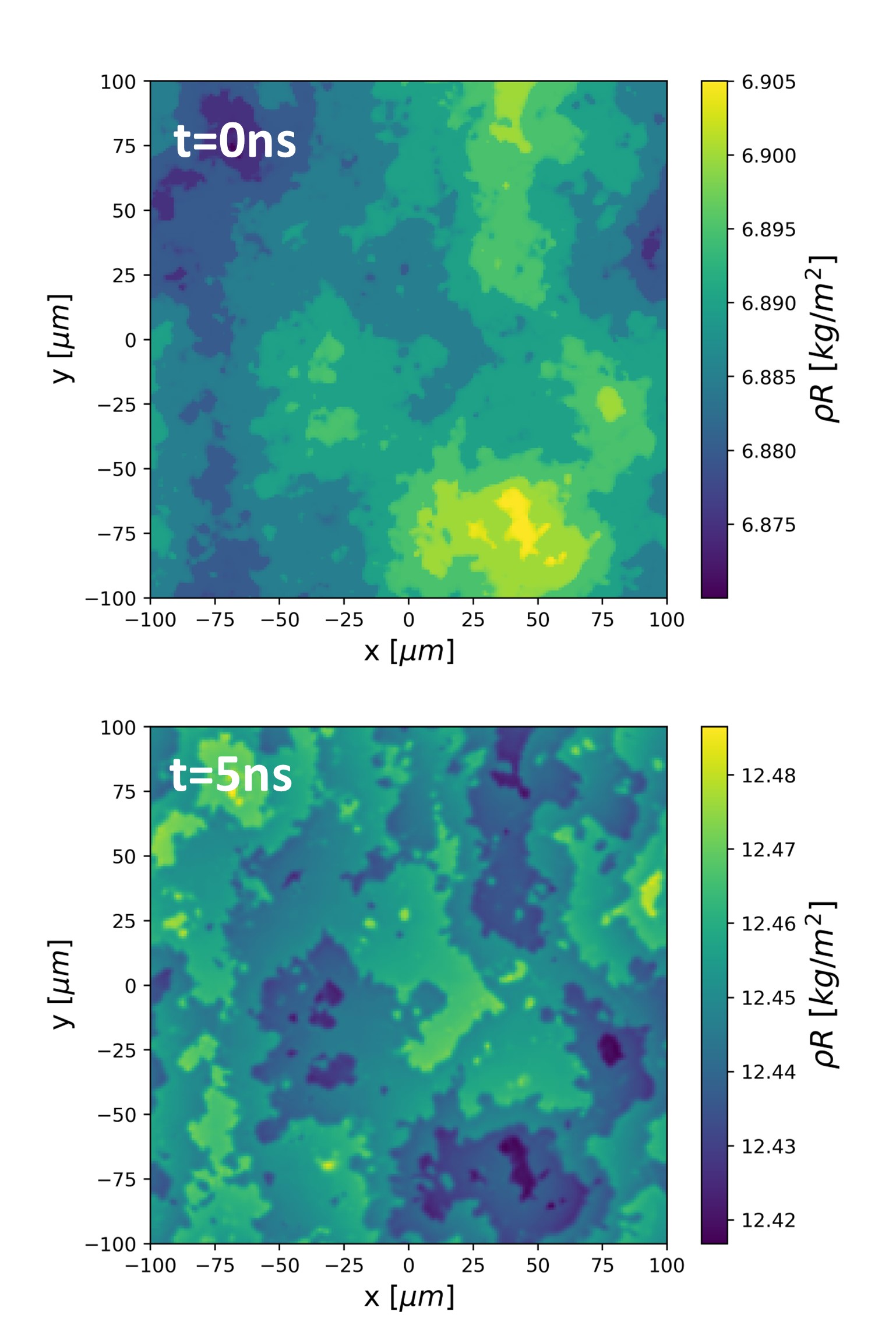} \caption{ \label{fig:rhoR_noB} Line-integrated density ($\rho R$) for the unmagnetized test case at t=0ns and t=5ns, showing inversion of the original perturbation. }	
\end{figure}
An initial perturbation is applied through varying the location of the ice-ablator interface in the direction of shock propagation. $10^4$ sinusoidal variations are applied with mode numbers in the range (0,100), giving an overall bubble-spike height of $\approx$3$\mu m$. Figure \ref{fig:rhoR_noB} shows the initial mass density integrated along the shock propagation direction ($\rho R$). The perturbation is periodic in $x$ and $y$, which is the same as the simulation boundary conditions.

A similar setup has been used to study the Rayleigh Taylor instability growth with self-generated \cite{walsh2023} and applied magnetic fields \cite{walsh2021magnetized}. The latter case found that 3-D simulations were essential to capture the evolution once a magnetic field was externally applied. 

This paper is organized as follows. Section \ref{sec:noB} demonstrates unmagnetized 3-D Richtmyer Meshkov growth as a reference for the subsequent sections. Section \ref{sec:high_temp} presents a case where the applied magnetic field suppresses growth of some modes; for this a high temperature is used at the ice-ablator interface. Section \ref{sec:low_temp} then uses a lower temperature more appropriate to ICF implosions; no significant impact of the applied field is found due to substantial resistive diffusion. Finally, section \ref{sec:postprocess} post-processes a 1-D HYDRA simulation of a NIF implosion to show that resistive diffusion is expected to be dominant in these systems. 

\section{Unmagnetized Richtmyer Meshkov Growth \label{sec:noB}}

This section gives a background on the Richtmyer Meshkov test problem utilized by studying the perturbation growth without an applied magnetic field. 

Figure \ref{fig:rhoR_noB} shows the amplification of integrated mass density ($\rho R$) over the 5ns after the shock has transited past the interface. Note that the initial perturbation has become inverted; in other words, spikes have become bubbles and bubbles have become spikes. This is a classic feature of Richtmyer Meshkov \cite{zhou2017}. The overall bubble-spike height, which started at around 3$\mu$m, is amplified to 6$\mu$m.

\begin{figure}
	\centering
	\centering
	\includegraphics[scale=0.6]{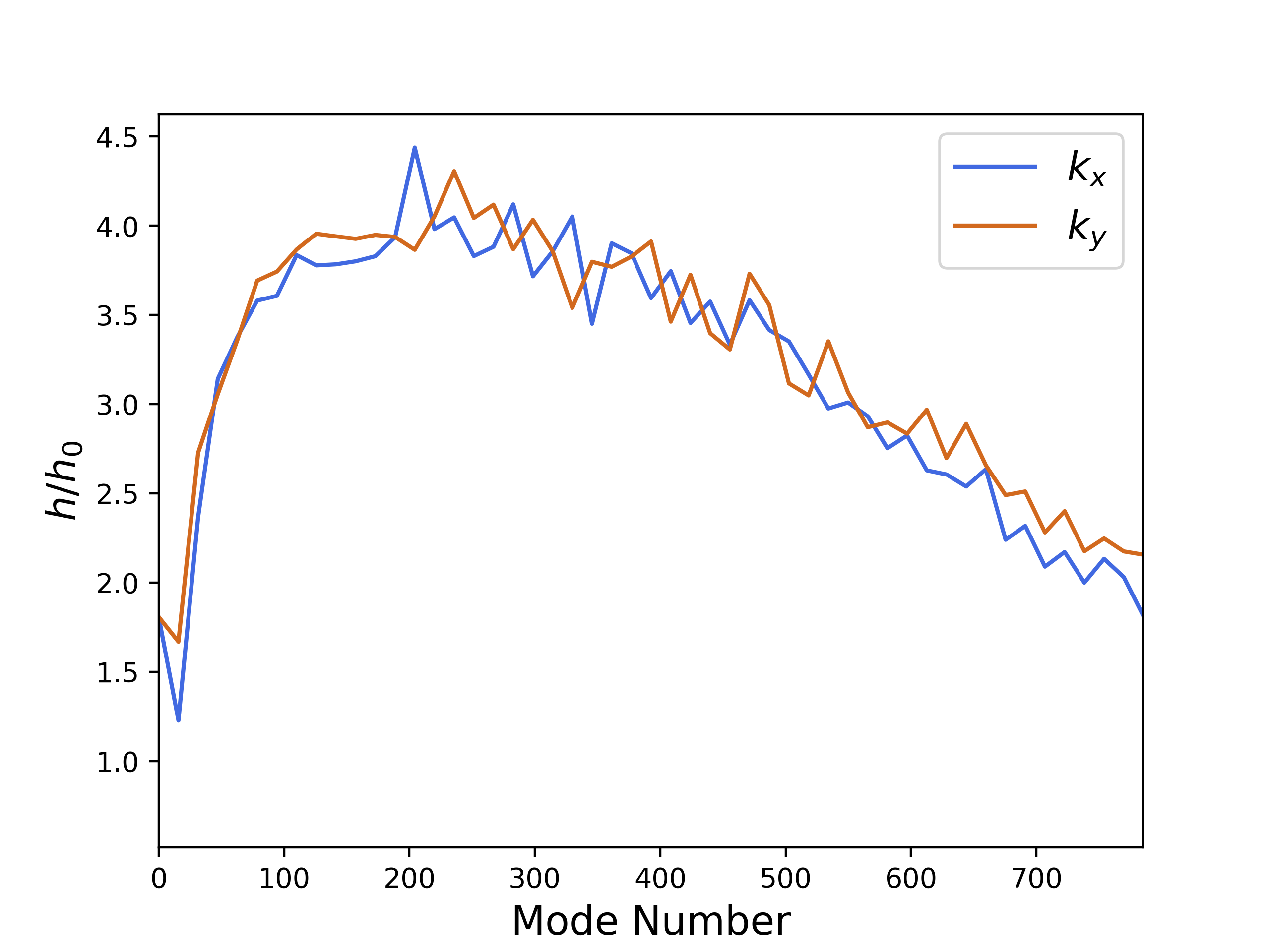} \caption{ \label{fig:amp_vs_k_noB} Amplification of the unmagnetized perturbation modes by the Richtmyer-Meshkov instability between t=0ns and t=5ns (from figure \ref{fig:rhoR_noB}.) }	
\end{figure}

The amplification for individual modes is quantified by taking the Fourier transforms of $\rho R$ maps from figure \ref{fig:rhoR_noB}. The growth of modes in $x$ and $y$ are kept separate, as later sections will show that magnetic fields can change the growth of modes along the applied field direction ($k_x$). Figure \ref{fig:amp_vs_k_noB} shows $k_x$ and $k_y$ for the unmagnetized case, with no discernible difference between the $x$ and $y$ directions, as expected. The peak amplification due to the Richtmyer Meshkov instability is found between modes 120-300, giving an amplification of $h/h_0 \approx$4.

\section{High Temperature Interface \label{sec:high_temp}}

\begin{figure}
	\centering
	\centering
	\includegraphics[scale=0.58]{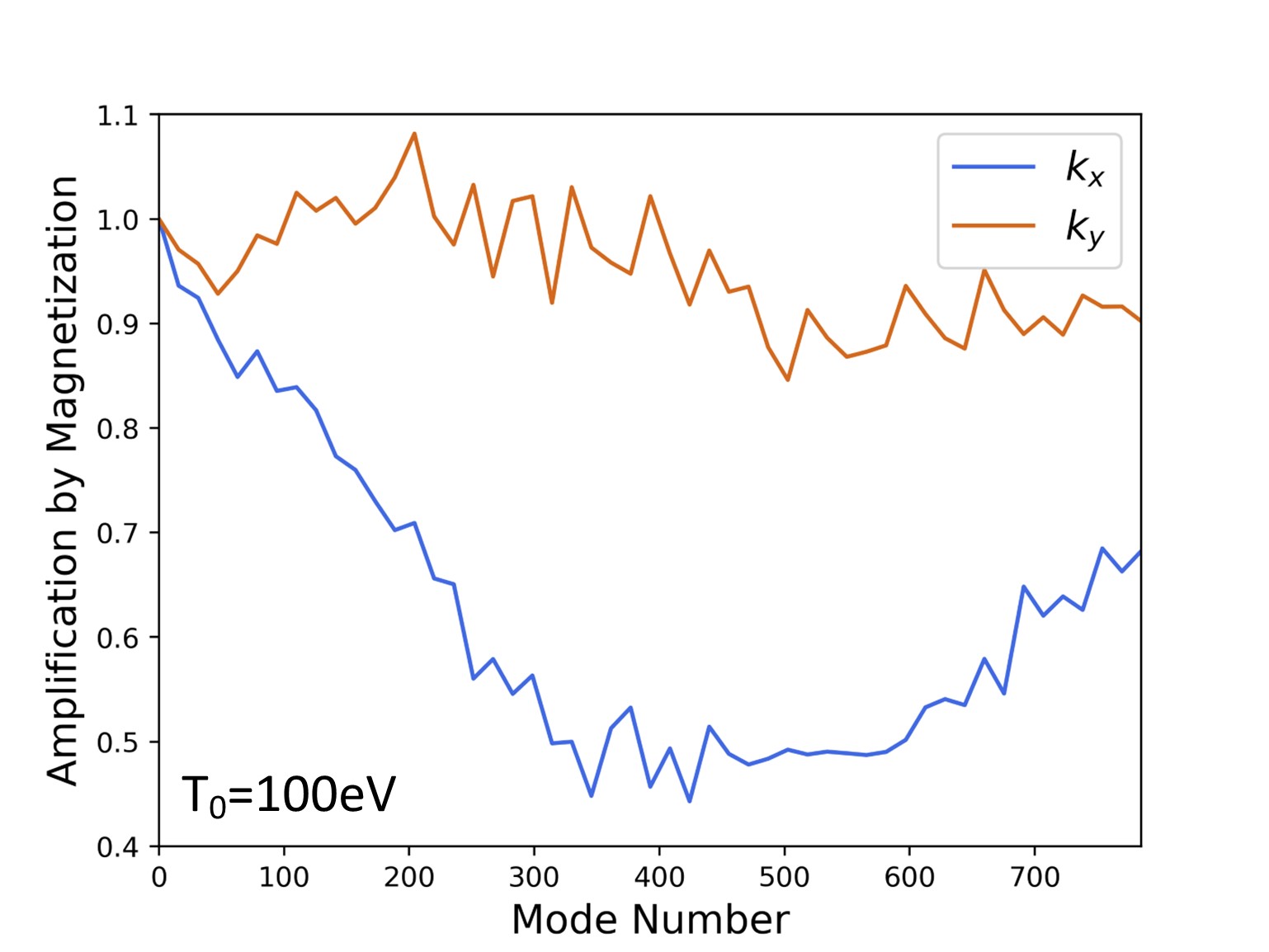} \caption{ \label{fig:amp_vs_k_100eV} Amplification of perturbation modes due to a 30T magnetic field at t=5ns for the test case using an initial 100eV interface temperature. }	
\end{figure}

This section presents results for an interface temperature of 100eV. 
\begin{figure*}
	\centering
	\centering
	\includegraphics[scale=0.5]{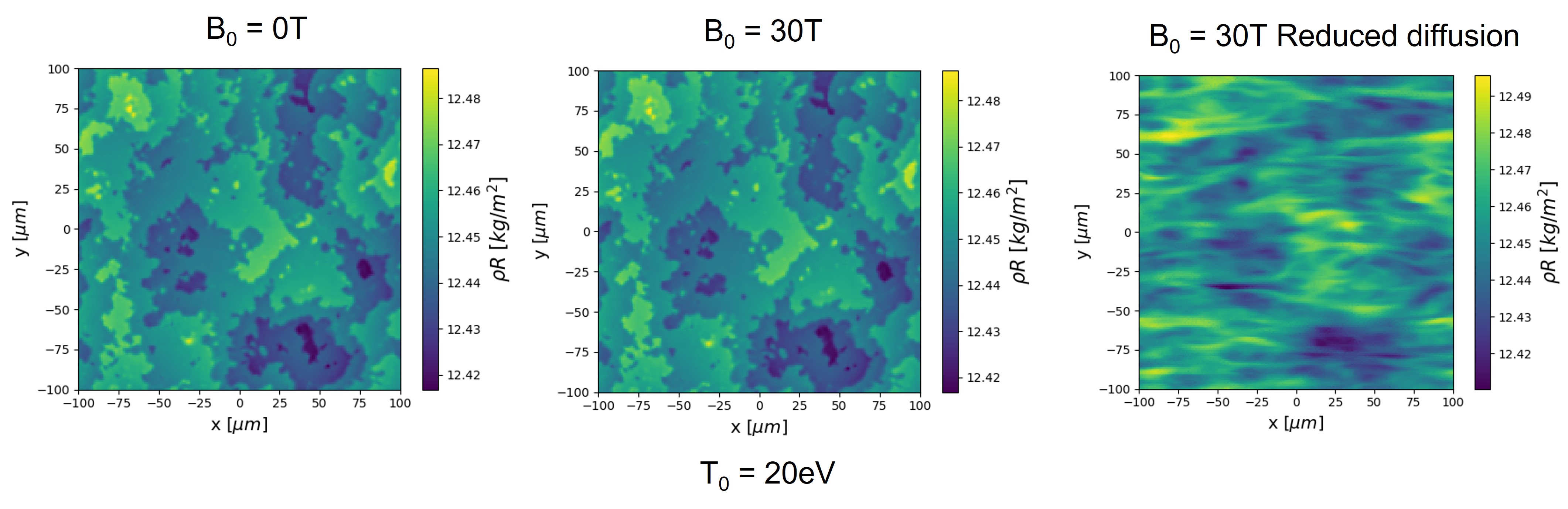} \caption{ \label{fig:rhoR_20eV} Line-integrated density ($\rho R$) at t=5ns for cases using a 20eV interface temperature. The first case is unmagnetized, the second uses a 30T transverse applied field, and the third uses the same magnetic field but reducing the diffusivity of the plasma by 5 orders of magnitude.}	
\end{figure*}

Figure \ref{fig:amp_vs_k_100eV} shows the impact of a 30T applied magnetic field on the modes along the magnetic field ($k_x$) and perpendicular to the magnetic field ($k_y$). The perturbations along the applied field direction are damped by the tension force in the magnetic field lines; as the perturbations grow they must do work on the magnetic field. This results in an approximate 50\% suppression of modes between 300-600. Note that these modes are shorter scale-length than the modes that undergo the most growth (from figure \ref{fig:amp_vs_k_noB}); this is due to the strong dependence of magnetic stabilization on the wavelength, as shown by equation \ref{eq:Bstab}. This behavior is similar to what has been found for the magnetized Rayleigh-Taylor instability \cite{walsh2021magnetized}.

The modes perpendicular to the magnetic field are largely unchanged by magnetization, which is in-line with previous 3-D simulations of perturbation growth with pre-imposed magnetic fields\cite{stone2007,walsh2021magnetized}. Quite simply, the perturbations in $y$ are able to grow without bending the field lines.

While there is substantial suppression at low modes, the magnetic field has no substantial impact above mode 1000. This is likely because the test problem does not have substantial mode growth at these small length-scales (as seen in figure \ref{fig:amp_vs_k_noB}). If the perturbations do not grow and perturb the field lines, there will be no magnetic tension associated with these modes to stabilize.

\section{Low Temperature Interface \label{sec:low_temp}}

\begin{figure}
	\centering
	\centering
	\includegraphics[scale=0.7]{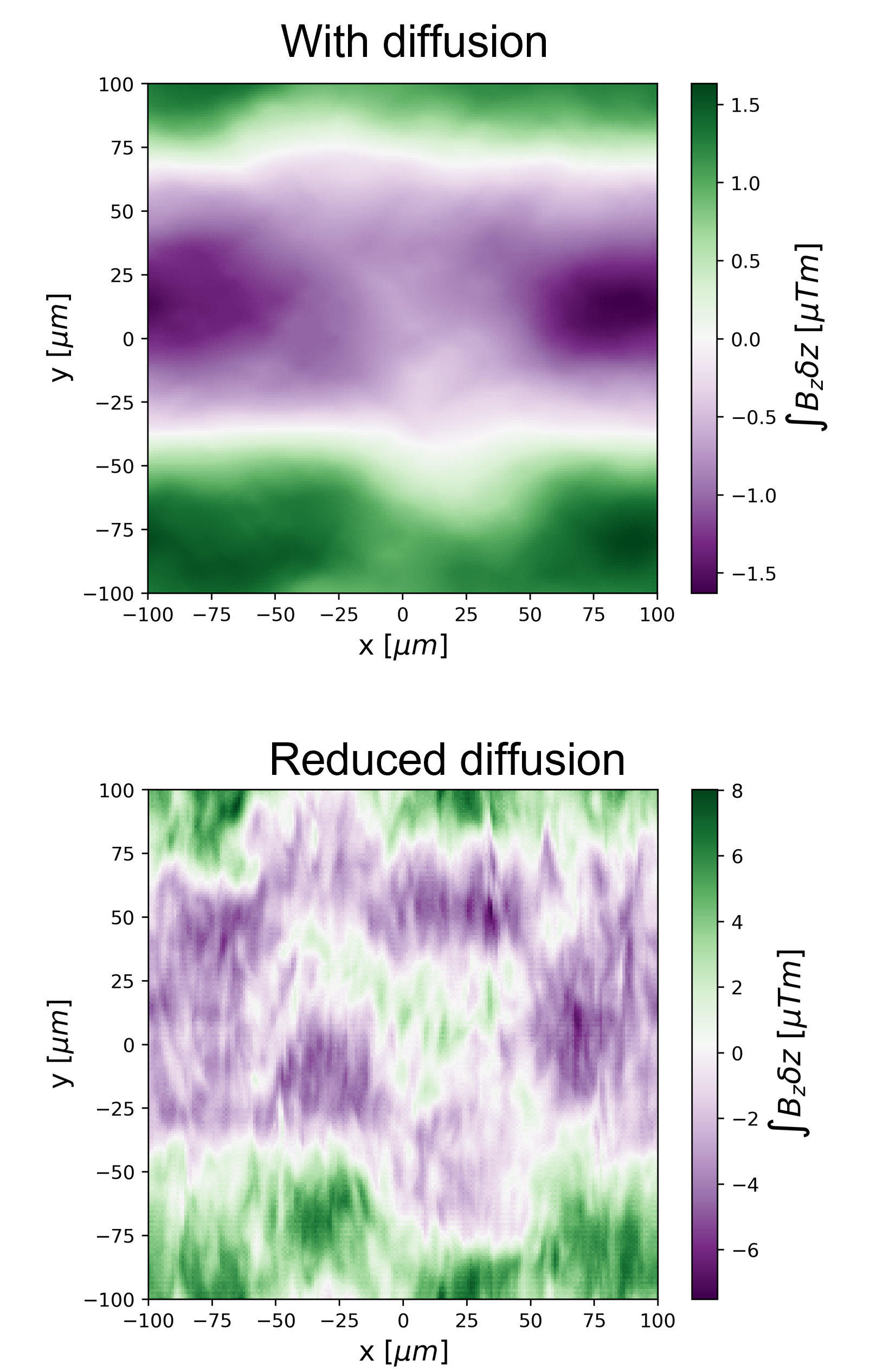} \caption{ \label{fig:intBz} The line-integrated $z$-component of the magnetic field (which is a measure of how twisted the magnetic field has become) for $B_0 = 30T$ and an interface temperature of 20eV. The second plot uses a reduced magnetic diffusivity by 5 orders of magnitude, showing much greater structure on small scales, which suppresses the high modes. }	
\end{figure}

For this section, the temperature at the ice-ablator interface is lowered to 20eV in order to demonstrate the role of resistive diffusion in ICF implosions.

Figure \ref{fig:rhoR_20eV} shows the unmagnetized (left) and magnetized (center) $\rho R$ maps at 5ns, with no discernible difference between the two. As the perturbations grow, the magnetic field is not twisted with the bulk plasma; instead the magnetic field diffuses. 

The final $\rho R$ plot in figure \ref{fig:rhoR_20eV} shows the perturbation for a 30T simulation where the resistive diffusivity in the plasma has been suppressed by 5 orders of magnitude, in order to demonstrate the significance of this effect. With reduced diffusivity, the $\rho R$ map develops clear striations along the applied field direction, with substantial suppression of modes in this direction.

To further show the impact of resistivity, a metric for magnetic field bending in the plasma is formulated. As the perturbations grow predominantly in the $z$ direction, the magnetic field component $B_z$ is integrated along $z$. For no perturbation, $B_z=0$ at all times. 

Figure \ref{fig:intBz} shows $\int B_z \delta z$ for the cases using the nominal resistive diffusivity and the reduced diffusivity. Without diffusion the magnitude of the twisting is clearly higher, but also the magnetic field has a much greater high-mode component. It is the high-mode component that is expected to be dominant in suppressing the perturbation growth, as the magnetic tension scales with $1/\lambda^2$.

\section{Magnetic Reynolds Number in ICF Implosions \label{sec:postprocess}}

\begin{figure}
	\centering
	\centering
	\includegraphics[scale=0.65]{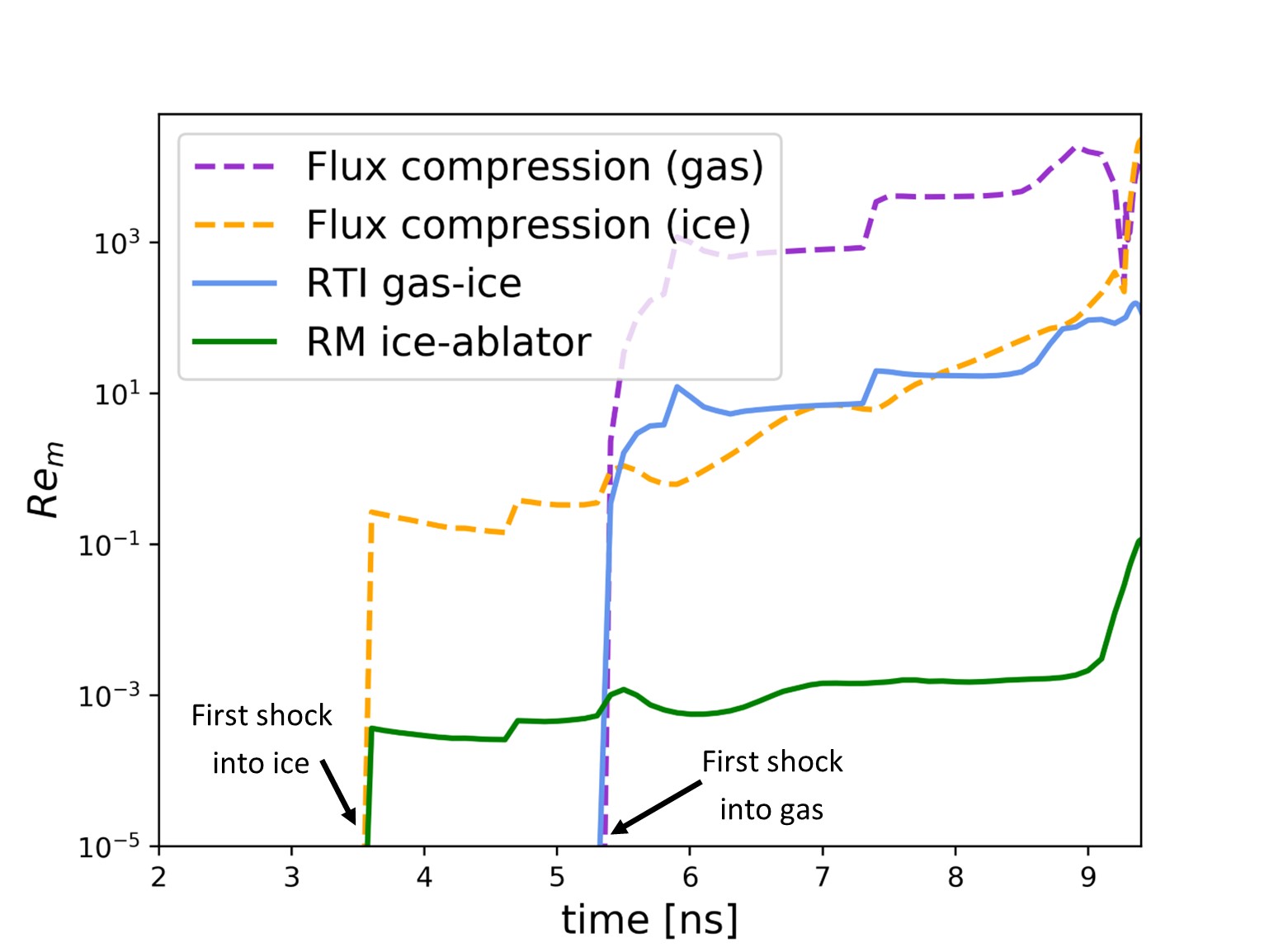} \caption{ \label{fig:Rem} Magnetic Reynolds number of the Richtmyer-Meshkov instability at the ice-ablator interface and the Rayleigh-Taylor instability at the gas-ice interface for a 1-D HYDRA simulation of N210808. }	
\end{figure}

So far 2 cases have been considered: one where the magnetic field is bent as the perturbation grows and subsequently stabilizes the interface; another where the resistive diffusion prevents the small-scale twisting of magnetic field. This section uses the magnetic Reynolds number as a metric for diffusivity and post-processes a 1-D HYDRA simulation to assess where and when resistivity is important in ICF implosions.

The magnetic Reynolds number is defined as:

\begin{equation}
	Re_m = \frac{U L}{\eta}
\end{equation}
Where $U$ is some velocity, $L$ some length-scale and $\eta$ is the resistive diffusivity. It can be assumed that resistivity is negligible for $Re_m \gg 1$.

ICF implosions are typically considered to be conductive. Indeed, a number of experimental\cite{knauer2010,gotchev2009,lewis2023,schmit2014} and computational\cite{davies2015,walsh2022} studies have found that the magnetic flux is highly compressed during implosion. For an ICF implosion, $U$ is taken as the peak implosion velocity and $L$ as the capsule radius\cite{hohenberger2012}:

\begin{equation}
	Re_{m,flux comp.} = \frac{V_{imp} R_{cap}}{\eta}
\end{equation}
Even better, it is possible to track the plasma properties at the different interfaces throughout the implosion. The magnetic diffusivitiy is independent of density\cite{epperlein1986}, so only the velocity, electron temperature and radius are needed. 

A 1-D HYDRA\cite{marinak1996} simulation of a NIF indirect-drive implosion is post-processed to estimate the impact of applied magnetic fields at different interfaces. The design chosen is for N210808, which was the first experiment to reach scientific ignition \cite{abu-shwareb2022}. The peak neutron production for this implosion was at t=9.28ns. 

Of course, the plasma properties vary either side of each interface, so decisions must be made in order to quantify the properties at the interface. The compressed magnetic field in the gas diffuses outwards into the ice, so plasma properties on the gas side of the gas-ice interface are used. Likewise, flux compression in the ice diffuses outwards into the ablator, so the properties of the ice adjacent to the ablator are used. The length-scale for the gas is taken as the time-dependent gas radius. The length-scale for the ice is taken as the ice thickness, which begins at 100$\mu$m. 

The flux compression magnetic Reynolds numbers are plotted in figure \ref{fig:Rem}. The times when the first shock passes into the ice and gas are indicated. The gas instantly becomes conductive, with $Re_{m,flux comp.}>10$ after the first shock passes through. For much of the implosion $Re_{m,flux comp.}>1000$, which shows that resistive diffusion has little impact on the flux compression. 

1.9ns after the first shock passes into the ice, the magnetic Reynolds for the ice exceeds 1; this is primarily because the temperature of the ice-ablator interface is low (between 4eV and 10eV at this time). Note that the ice only moves 50$\mu$m during the 1.9ns, so the marginal diffusivity may not be important. While integrated simulations of magnetized layered implosions do not report significant resistive diffusion in the ice \cite{perkins2017,perkins2013,walsh2019,walsh2022}, the calculations here suggest it is a process worth assessing during the design of implosions.  

Assessing the magnetic diffusivity for perturbation growth requires the use of the perturbation wavelength and growth rate, both of which are much smaller than used for estimating bulk magnetic flux compression. Here is the form for the magnetic Reynolds relevant to perturbation growth:

\begin{equation}
	Re_{m,pert.} = \frac{\frac{\partial h}{\partial t} \lambda}{\eta}
\end{equation}

For the 100eV interface temperature used in section \ref{sec:high_temp}, $Re_{m,pert} = 0.1$ for mode 30 (the most highly suppressed mode). This shows that $Re_{m,pert}$ does not need to exceed 1 for magnetic tension to stabilize perturbation growth. For the 20eV interface in section \ref{sec:low_temp}, $Re_{m,pert} \approx 0.01$ for mode 30, which saw no impact from the magnetic field. Therefore, it can be assumed that for $Re_{m,pert} \le 0.01$ the magnetic field does not stabilize perturbations.

Figure \ref{fig:Rem} plots the estimated magnetic Reynolds number at the ice-ablator interface throughout the implosion. A growth rate $\partial h/\partial t =$ 1$\mu$m/ns is assumed, which is taken from more detailed Richtmyer-Meshkov estimates \cite{weber2023}. A mode number of 2000 is used, as this is thought to have the most detrimental growth \cite{weber2023}.  Before 5.5ns $Re_{m,pert}<10^{-3}$, as the temperature is very low ($<$10eV). While the interface is progressively heated from 7ns onwards (reaching 20eV by 8.1ns), the capsule radius is simultaneously reduced, which lowers the perturbation length-scales.

Much of the Richtmyer-Meshkov growth is thought to occur in the earlier phase of the implosion, long before the ice-ablator interface reaches temperatures of 20eV. Even mode 200 perturbations will still have $Re_{m,pert}<10^{-2}$.

The gas-ice interface, on the other hand, is expected to have much lower resistivity; the temperature is much higher ($>$100eV from 6ns) and the dominant modes are much lower (figure \ref{fig:Rem} uses mode 40 \cite{weber2015,walsh2021a}). In addition, the perturbation growth is much stronger (here $\partial h/\partial t = 10\mu$m/ns). Therefore, it is no surprise that simulations including resistivity have found significant magnetic stabilization of the Rayleigh-Taylor instability at this interface \cite{perkins2017,walsh2019,walsh2021magnetized}.

Note that the calculations made here use a Spitzer resistivity \cite{epperlein1986}. Future work should assess uncertainty in the magnetic Reynolds by using other resistivity models, e.g. those given by Lee-More \cite{lee1984}.

\section{Conclusion}

A 3-D magnetized Richtmyer-Meshkov test problem has been used to investigate magnetic stabilization of perturbations at the ice-ablator interface of ICF implosions. Resistive diffusion is found to be a dominant magnetic transport process, preventing the magnetic field from bending with the perturbation growth. This is unlike simulation results for Rayleigh-Taylor growth at the hot-spot edge, which suggest strong stabilization \cite{walsh2021magnetized}.

While the high yield N210808 design was post-processed and found to be highly resistive at the ice-ablator interface, it remains possible that other designs would be more conductive and therefore benefit more strongly from magnetic stabilization. However, an implosion with $\approx 100eV$ interface temperatures would already be a high adiabat design and would be unlikely to undergo high compression.

Nonetheless, the high temperature interface simulated saw 50\% stabilization for modes 300-600. A few important caveats should be stated explicitly here. First, the magnetic field is found to only stabilize the modes along the magnetic field lines; i.e. the overall bubble-spike height is not reduced by magnetization, although the interface surface area is reduced. Secondly, for a spherical implosion the magnetic field is only in the plane of the ablator at the capsule waist; therefore the magnetic field would only improve compressibility at the waist and not at the capsule poles. 

While these results suggest magnetic fields will not improve compressibility at the ice-ablator interface, magnetic stabilization of the Rayleigh-Taylor instability at the hot-spot edge is still an exciting area of research. Experiments in the coming years will attempt to demonstrate these effects.

Dedicated shock tube experiments attempted to measure suppression of perturbation growth due to a transverse magnetic field \cite{manuel2021}. As in the work in this paper, the inclusion of resistivity resulting in the smearing of magnetic field profile, which reduced magnetic stabilization.

\section*{Acknowledgements}
This work was performed under the auspices of the U.S. Department of Energy by Lawrence Livermore National Laboratory under Contract DE-AC52-07NA27344. 

Work supported by LLNL LDRD projects 23-ERD-025 and 20-SI-002.

This document was prepared as an account of work sponsored by an agency of the United States government. Neither the United States government nor Lawrence Livermore National Security, LLC, nor any of their employees makes any warranty, expressed or implied, or assumes any legal liability or responsibility for the accuracy, completeness, or usefulness of any information, apparatus, product, or process disclosed, or represents that its use would not infringe privately owned rights. Reference herein to any specific commercial product, process, or service by trade name, trademark, manufacturer, or otherwise does not necessarily constitute or imply its endorsement, recommendation, or favoring by the United States government or Lawrence Livermore National Security, LLC. The views and opinions of authors expressed herein do not necessarily state or reflect those of the United States government or Lawrence Livermore National Security, LLC, and shall not be used for advertising or product endorsement purposes.
	\section*{Bibliography}
	
	\bibliographystyle{unsrt}
	\bibliography{library}

\end{document}